SPACE SCIENCES

# Evidence for quasar fast outflows being accelerated at the scale of tens of parsecs

Zhicheng He[1,2]*, Guilin Liu[1,2]*, Tinggui Wang[1,2]*, Guobin Mou[3]*, Richard Green[4], Weihao Bian[5], Huiyuan Wang[1,2], Luis C. Ho[6,7], Mouyuan Sun[8], Lu Shen[1,2], Nahum Arav[9], Chen Chen[10], Qingwen Wu[11], Hengxiao Guo[12], Zesen Lin[1,2], Junyao Li[1,2], Weimin Yi[13]

Quasar outflows may play a crucial role in regulating the host galaxy, although the spatial scale of quasar outflows remain a major enigma, with their acceleration mechanism poorly understood. The kinematic information of outflow is the key to understanding its origin and acceleration mechanism. Here, we report the galactocentric distances of different outflow components for both a sample and an individual quasar. We find that the outflow distance increases with velocity, with a typical value from several parsecs to more than one hundred parsecs, providing direct evidence for an acceleration happening at a scale of the order of 10 parsecs. These outflows carry ~1% of the total quasar energy, while their kinematics are consistent with a dust-driven model with a launching radius comparable to the scale of a dusty torus, indicating that the coupling between dust and quasar radiation may produce powerful feedback that is crucial to galaxy evolution.

## INTRODUCTION

Blueshifted broad absorption lines (BALs) are observed in the spectra of 10 to 40% (1, 2) of quasars with the intrinsic BAL fraction at $z \sim 2.5$ that is potentially as high as 80% (2), implying that outflows, of which even the fastest (up to $0.1c$) ones, are not rare. Some theoretical models consider such outflows to be a possible agent of the so-called active galactic nucleus (AGN)/quasar feedback, which may play a role in regulating the growth of supermassive black holes (SMBHs) and the evolution of their host galaxies (3–6). For other possible feedback mechanisms, see the introduction of (7). The galactocentric distance ($R$) and kinematics of the outflows are critical for understanding their origin and influence on the surroundings at (inter-) galactic scales.

During the past two decades, efforts have been made to constrain the galactocentric distances of BAL outflows (8–13). These findings, sometimes controversial and debatable, range from parsecs to kiloparsecs. Other outflow parameters, e.g., those delivering kinematic information, remain largely uncertain as well, rendering the understanding of the origin and acceleration mechanism a long-standing challenge. The velocity profile, i.e., the velocity as a function of galactocentric distance $R$, is fundamental for characterizing the outflow kinematics. Although the spatially resolved kinematics of the line-emitting gas around AGN on scales of 0.1 to a few kiloparsecs have been reported in a large amount of literature [e.g., see (14–19) for a review], the velocity profiles of outflows remain largely unexplored on the observational front, especially those of fast BAL outflows.

Conventional models for subrelativistic outflows often invoke winds originating from the accretion disk of the central black hole (20, 21), suggesting an acceleration phase taking place at scales substantially smaller than a parsec. Models with a larger launching radius exist as well, and a dusty outflow driven by radiation pressure has long been discussed (22–25). This scenario naturally predicts a reddened ultraviolet (UV)/optical continuum of BAL quasars, in comparison with that of the non-BAL quasars. In line with this, BAL quasars do appear redder than their non-BAL counterparts in many sample analyses (26–29). In particular, for a typical SMBH with $M_{BH} = 10^9 M_\odot$, the disk wind it drives is accelerated at the scale of 0.01 to 0.1 pc, while in case of a dusty outflow, the acceleration occurs at locations beyond the inner radius of the dusty torus (~1 pc). Therefore, determining the spatial scale for the acceleration phase of the outflowing gas is the key to discriminate between these two types of models.

BAL variability proves to be a novel and powerful probe of the nature of quasar outflows. BAL troughs vary on time scales ranging from days to years (30–34). There are two possible origins of BAL variability: the tangential movement of the absorbing gas or change in the ionizing radiation incident on the gas. Studies of individual objects show evidence for both origins, with some supporting the scenario of moving absorbing gas [e.g., (9, 35)] and others supporting the scenario of varying ionization state [e.g., (36–38)]. Large sample studies indicate that the majority of BAL variability is driven by variation of the ionizing continuum (32, 39). We further proved that at least 80% of BAL variations are driven by the variation of ionizing continuum (40). On the basis of the above work, we focus on BAL variability driven by ionizing continuum variation. The ionization state of a gaseous outflow needs a period of time to respond to changes in the ionizing continuum [the recombination time scale, $t_r$ (39, 41–43)]. Therefore, the gas ionization state is connected to the intensity of the ionizing continuum over $t_r$.

[1]CAS Key Laboratory for Research in Galaxies and Cosmology, Department of Astronomy, University of Science and Technology of China, Hefei, Anhui 230026, China. [2]School of Astronomy and Space Science, University of Science and Technology of China, Hefei 230026, China. [3]School of Physics and Technology, Wuhan University, Wuhan 430072, China. [4]Steward Observatory, University of Arizona, Tucson, AZ 85721-0065, USA. [5]Department of Physics and Institute of Theoretical Physics, Nanjing Normal University, Nanjing 210023, China. [6]Kavli Institute for Astronomy and Astrophysics, Peking University, Beijing 100871, China. [7]Department of Astronomy, School of Physics, Peking University, Beijing 100871, China. [8]Department of Astronomy, Xiamen University, Xiamen, Fujian 361005, China. [9]Department of Physics, Virginia Tech, Blacksburg, VA 24061, USA. [10]School of Physics and Astronomy, Sun Yat-sen University, Zhuhai 519082, China. [11]School of Physics, Huazhong University of Science and Technology, Wuhan 430074, China. [12]Department of Physics and Astronomy, University of California, Irvine, 4129 Frederick Reines Hall, Irvine, CA 92697-4575, USA. [13]Department of Astronomy and Astrophysics, The Pennsylvania State University, 525 Davey Lab, University Park, PA 16802, USA.
*Corresponding author. Email: zcho@ustc.edu.cn (Z.H.); glliu@ustc.edu.cn (G.L.); twang@ustc.edu.cn (T.W.); gbmou@ustc.edu.cn (G.M.)

In our previous work (*43*), we statistically analyzed a sample of 915 Sloan Digital Sky Survey data release 14 (SDSS DR14) quasars, where the distribution of $t_r$ of BAL outflows was derived, along with that of the galactocentric radius $R$, which is found to be typically several to tens of parsecs. However, the approach in (*43*) treated the BAL trough as a single component and overlooked the outflow kinematics. In reality, BAL absorbers frequently consist of a number of components with different velocities, column densities, and ionization parameters (*44*) that are blended along our line of sight. Here, we report a new method applicable to both individual quasars/components and a sample, large or small. Using this method, we derive the characteristic values of $t_r$ and, hence, the distances $R$ for different velocity components of the outflowing gas so that some kinematic information about these winds can be obtained.

## RESULTS

As a heuristic explanation of this new method, we suppose that there exist three observations for a BAL quasar, which satisfy the following requirements (see Methods and fig. S1 for details). First, the time interval between the observational epochs 1 and 2, $\Delta T_{12}$, is shorter than the time interval between the epochs 1 and 3, $\Delta T_{13}$, i.e., $\Delta T_{12} < \Delta T_{13}$. Second, the amplitude of continuum flux variation at 1500 Å for epochs 1 and 2 is greater than that for epochs 1 and 3, i.e., $|(\Delta L/L)_{12}| > |(\Delta L/L)_{13}|$. If the recombination time scale $t_r$ of the outflowing gas is relatively short, e.g., $t_r < \Delta T_{12}$, we would expect that the amplitude of BAL variation between epochs 1 and 2 is greater than that of epoch 1 and 3; we mark this case as "G1." If the recombination time scale $t_r$ is relatively long, e.g., $t_r > \Delta T_{12}$, then there will be no significant BAL variation between epochs 1 and 2. As a result, we will find that the amplitude of BAL variation between epochs 1 and 2 is smaller than that of epoch 1 and 3; we mark this case as "G2." In general, the probability of the amplitude of BAL variation between epochs 1 and 2 being larger than that of epochs 1 and 3 gradually decreases for an increasing $t_r$. The G1 and G2 certification process are shown in fig. S2, and an example of the G1 and G2 cases is shown in fig. S3. If the number counts of G1 and G2 certificated from the observational data are $N_{G1}$ and $N_{G2}$, respectively, we define the fraction of G1 as: $F_{G1} = N_{G1}/(N_{G1} + N_{G2})$. The G1 fraction, $F_{G1}$, will decrease when $t_r$ increases. In this work, we use the $F_{G1}$ curve (see Method and Fig. 1) measured from the SDSS DR16 to constrain $t_r$.

Outflow features are often embedded in the absorption line spectra in a complex manner, and the outflowing material lying in our line of sight frequently produces multiple components. To optimize the statistical significance of our kinematic analyses, we categorize the sample C IV doublet (1548 Å, 1550 Å) BAL troughs into three velocity bins, namely, low (0 to 5000 km s$^{-1}$), medium (5000 to 10,000 km s$^{-1}$), and high velocity (>10,000 km s$^{-1}$). As shown in Fig. 1A, the G1 fraction $F_{G1}(t_r)$ curve is measured from the SDSS DR16 sample (see Method and table S1), which contains 46 quasars with $1.9 < z < 4.7$ and an average bolometric luminosity $L_{bol} = 10^{46.5}$ erg s$^{-1}$. The composite spectrum of the sample is shown in fig. S4. The G1 fractions are 67.9 ± 4.0 % (93/137), 57.7 ± 5.7 % (79/137), and 40.9 ± 4.2 % (56/137) for low-, medium- and high-velocity bins, respectively (note that the uncertainties are estimated from binomial distributions, and the significance of the difference between low- and high-velocity bins is about 4.7σ). Correspondingly, $t_r$ is $10^{0.56 \pm 0.19}$, $10^{1.00 \pm 0.21}$, and $10^{2.35 \pm 0.41}$ days for low-, medium-, and high-velocity outflows, respectively, showing a clear trend of $t_r$ increasing with the outflow velocity (Fig. 1B). We should note that the signal-to-noise ratio (S/N) of absorption features will generally be lower at higher velocities. As a result, the G1 fraction, as well as the recombination time scale, will be affected by the S/N. If the intrinsic G1 fraction is larger (or smaller) than 50%, the low S/N will reduce (or increase) the G1 fraction and make it closer to 50%. Therefore, the intrinsic G1 fractions of low and medium velocity should be larger than 67.9 and 57.7%, respectively, and that of the high velocity should be smaller than 40.9%. In view of this, the difference of intrinsic G1 fractions between the high-velocity bin and the other bins should be more significant.

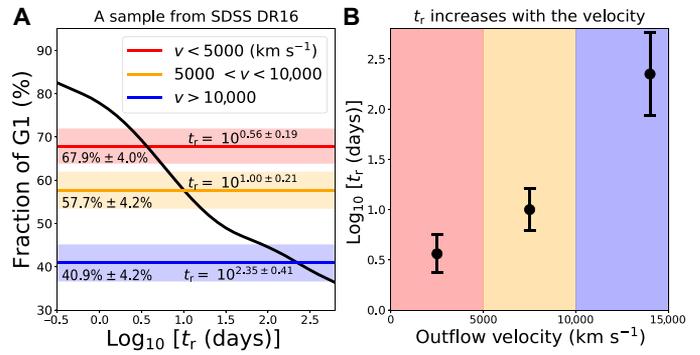

**Fig. 1. The recombination time scales of different velocity components of outflow gas for an SDSS sample.** The black curve represents the simulated G1 fraction curve (see Methods for details). The colored horizontal lines and corresponding shaded regions are the observed G1 fraction and the corresponding 1σ error. (**A**) The G1 fractions are 67.9 ± 4.0 % ( of 137), 57.7 ± 5.7 % ( of 137), and 40.9 ± 4.2 % (56 of 137) for the the low- (0 to 5000 km s$^{-1}$), medium- (5000 to 10,000 km s$^{-1}$), and high-velocity (>10,000 km s$^{-1}$) regions, respectively (the significance of the difference between low- and high-velocity bins is about 4.7σ). The corresponding recombination time scales $t_r$ are shown in (**B**). The vertical error bars mark the 1σ uncertainty. The sample shows a clear trend of $t_r$ increasing with the outflow velocity.

In the multi-epoch spectra of a quasar, SDSS J141955.26+522741.1 (hereafter J1419), there are five different velocity components (from A to E) as per our visual inspection (see Methods and fig. S5). The BAL variations of J1419 have been found to be driven by the variation of the ionizing continuum (see Methods and fig. S6). As a result, the above method of measuring $t_r$ can be applied to J1419. As shown in Fig. 2A and table S2, the G1 fractions are 64.3 ± 0.8 % (of 3980), 71.2 ± 0.7 % (of 3766), 60.3 ± 1.5 % (of 1033), 56.2 ± 1.3 % (793 of 1412), and 40.3 ± 1.9 % (258 640) for the troughs A to E, respectively. The G1 difference between the B and E components shows a significance of 15.5σ. Consequently, $t_r$ is $10^{0.90 \pm 0.03}$, $10^{0.67 \pm 0.02}$, $10^{1.06 \pm 0.07}$, $10^{1.27 \pm 0.08}$, and $10^{2.54 \pm 0.13}$ days for outflow components A to E (Fig. 2B), respectively. Likewise, this object also shows a clear trend of $t_r$ increasing with the outflow velocity.

The recombination time $t_r$ links the ionization state, the electron density, and the recombination rate (see Methods); meanwhile, the definition of the ionization parameter, $U = Q_H/(4\pi R^2 n_H c)$, demonstrates that the outflow distance $R$ to the central SMBH can be determined as long as the ionization state and electron density are pinned down or well constrained. Wang *et al.* (*39*) found that C IV, Si IV, and N V respond negatively to an increasing ionization parameter and constrained the ionization parameter of most BAL outflows to $\log_{10} U \gtrsim 0$. Furthermore, on the basis of the composite spectra of BAL quasars from SDSS, Hamann *et al.* (*13*) concluded that the minimum ionization parameter is $\log_{10} U \gtrsim -0.5$. These results validate our strategy of restricting our consideration to the range of ionization state: $\log_{10} U = -0.5$, 0, and 1 and taking $\log_{10} U = 0$ as the fiducial value for deriving the final results (although we report the results on the basis of other ionization parameter values as well, for reference).

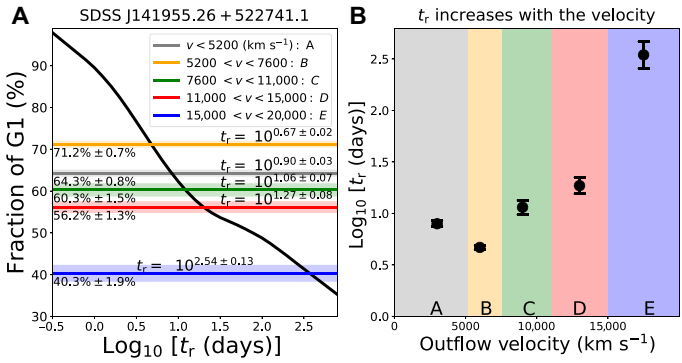

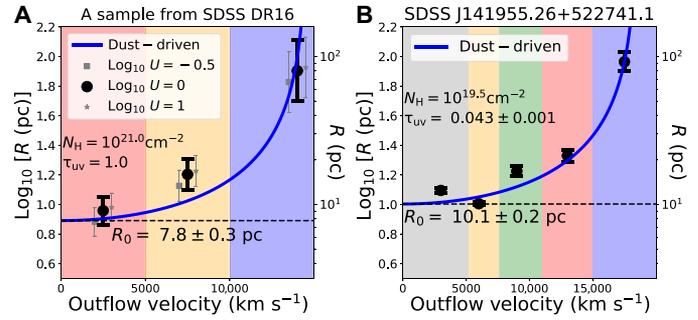

**Fig. 2. The recombination time scales of different velocity components of outflow gas of an individual quasar J1419.** (**A**) The G1 fractions are 64.3 ± 0.8 % (2559 of 3980), 71.2 ± 0.7 % (2682 of 3766), 60.3 ± 1.5 % (623 of 1033), 56.2 ± 1.3 % (793 of 1412), and 40.3 ± 1.9 % (258 of 640) for troughs A to E, respectively. The colored horizontal lines and corresponding shaded regions are the observed G1 fraction and the corresponding 1σ error. The significance of the G1 difference between components B and E is about 15.5σ. The corresponding recombination time scales $t_r$ are shown in (**B**). The vertical error bars mark the 1σ uncertainty. Both the sample (Fig. 1) and this individual quasar show a clear trend of $t_r$ increasing with the outflow velocity.

**Fig. 3. The observational evidences of the acceleration of BAL outflow on the scale of 10 parsecs and the dust-driven model.** (**A**) For the sample, the outflow distances $R$ is $10^{0.96 \pm 0.10}$, $10^{1.20 \pm 0.11}$, and $10^{1.90 \pm 0.21}$ pc at $\log_{10} U = 0$ for the low-velocity, medium-velocity, and high-velocity regions, respectively. (**B**) For J1419, $R$ is $10^{1.10 \pm 0.02}$, $10^{1.00 \pm 0.01}$, $10^{1.23 \pm 0.04}$, $10^{1.33 \pm 0.04}$, and $10^{1.97 \pm 0.07}$ pc for component A to E, respectively. The vertical error bars mark the 1σ uncertainty. Both the sample and the individual object reveal that the outflows are being accelerated at the 10-pc scale, which is far beyond the scale of the disk wind acceleration phase (20, 21). We use the dust-driven model (Eq. 1) with the optical depth of dust in the UV band $\tau_{uv} = 1$ to fit the observational data of the sample. The best-fit outflow launching radius $R_0 = 7.8 \pm 0.3$ pc. For J1419, we treat the $\tau_{uv}$ as a free parameter and the best-fit launching radius of $R_0 = 10.1 \pm 0.2$ pc.

According to Eq. 5, the logarithmic electron density $\log_{10}(n_e/\text{cm}^{-3})$ of the sample is 4.93 ± 0.19, 4.44 ± 0.21, and 3.04 ± 0.41 for the low-, medium-, and high-velocity bins at $\log_{10} U = 0$, respectively. Adopting the average bolometric luminosity $L_{bol} = 10^{46.5} \text{erg s}^{-1}$, $R$ is $10^{0.96 \pm 0.10}$, $10^{1.20 \pm 0.11}$, and $10^{1.90 \pm 0.21}$ pc at $\log_{10} U = 0$ for the low-, medium-, and high-velocity bins, respectively (Fig. 3A). As for J1419, the ionization parameters of its five components are roughly $\log_{10} U \simeq -2$ (fig. S6), and therefore, we simply adopt $\log_{10} U = -2$. More precise ionization parameters need to be measured in the future. However, because of $R \propto U^{-1/2}$ at a given electron density, the uncertainty of distance is smaller than that of ionization parameter. The electron density $\log_{10}(n_e/\text{cm}^{-3})$ is 6.06 ± 0.03, 6.24 ± 0.02, 5.80 ± 0.07, 5.59 ± 0.08, and 4.32 ± 0.13 for component A to E, respectively. Adopting the bolometric luminosity $L_{bol} = 10^{45.9} \text{erg s}^{-1}$, we find that $R$ is derived to be $10^{1.10 \pm 0.02}$, $10^{1.00 \pm 0.01}$, $10^{1.23 \pm 0.04}$, $10^{1.33 \pm 0.04}$, and $10^{1.97 \pm 0.07}$ pc for component A to E, respectively (Fig. 3B). Therefore, both our sample and case studies deliver observational evidence for an acceleration of outflows happening at a scale of the order of 10 pc, far beyond where disk wind acceleration occurs (20, 21).

## DISCUSSION

It should be noted that the current uncertainties of $R$ shown in the work are purely random measurement and statistical errors. The uncertainties in luminosity only result in an offset in the resultant $R$ of different velocity bins but will not affect the overall difference (and trend) therein. The primary uncertainties in $R$ result from estimating the ionization parameters. Using the fiducial parameter $\log_{10} U = 0$ for the sample is a conservative strategy. The composite spectrum of our sample (see fig. S4) demonstrates that the equivalent width (EW) ratio of Si IV to C IV BALs decreases as outflow velocity increases. The EW(Si IV)/EW(C IV) distributions for the 46 quasars are also shown in fig. S4D. The mean and SD of EW(Si IV)/EW(C IV) distribution are 0.43 ± 0.19, 0.42 ± 0.18, and 0.37 ± 0.14 for low, mid, and high velocity, respectively. Evidently in the comparison, no significant statistical difference exists between low and high velocities. To further explore this, we examine the EW(Si IV)/EW(C IV) in every object, finding it to be higher in the low velocities than in high velocities in about two-thirds (30 of 46) of the sample. The above analysis implies that the ionization parameter of the high-velocity regime is larger (or at least remains on a plateau) than that of the low-velocity regime (12). If this difference is taken into account, then the significance of the distance difference between the low- and high-velocity outflows will scale up. We also note that each quasar in the sample exhibits BAL variations in all of its three velocity regimes, ruling out the possibility that the positive correlation between outflow velocity and the galactocentric distance is a false correlation caused by a probable positive correlation between luminosity and BAL velocity.

As we have concluded, the spatial scale at which outflows are being accelerated is the order of 10-pc scale, beyond the inner radius of the dusty torus [$R_{inner} \simeq 1$pc for $L_{bol} = 10^{46.5} \text{erg s}^{-1}$, using the dust reverberation mapping (45) of K band].

Dust is intrinsic to outflows, as the combination of SDSS and Widefield Infrared Survey Explorer (WISE) datasets has demonstrated (46). On the basis of the SDSS quasar catalog, Gibson et al. (1), Allen et al. (2), and Baskin et al. (28) find that, compared to non-BALQs, the median spectral energy distribution (SED) of broad absorption line quasars (BALQs) is consistent with being reddened from the extinction curve of the Small Magellanic Cloud (SMC) (47, 48) with $A_V = 0.06$ to 0.15 magnitude (mag). As for our sample, we find the non-BAL quasar template reddened by an SMC extinction curve with $A_V = 0.1$ to be remarkably consistent with the composite spectrum of the 46 BAL quasars (fig. S4A), validating our choice of adopting $A_V = 0.1$ hereafter in this work.

Both of the theoretical (*49*) and the observational [see figure 5 in (*48*)] SMC extinction curves peak around $\frac{1}{\lambda} \sim 10$ to $15\ \mu m^{-1}$ ($\lambda \sim 1000$ to 667 Å), where $A_\lambda$ is larger than $A_V$ by a factor of 10. For our sample quasars, we set the UV band extinction to be $10\ A_V$, corresponding to an optical depth of $\tau_{uv} \simeq A_{uv} = 10\ A_V \sim 1$. In a dust-driven outflow, the galactocentric distance $R$ and the outflow velocity are related analytically as below (see Methods for details)

$$R = \frac{1}{\frac{1}{R_0} - \frac{2\pi m_p N_H c v^2}{L_{uv}(1-e^{-\tau_{uv}})}} \quad (1)$$

where $v < v_{max} = \sqrt{\frac{L_{uv}(1-e^{-\tau_{uv}})}{2\pi R_0 m_p N_H c}}$, the maximum outflow velocity, $m_p = 1.67 \times 10^{-27}$ kg is the mass of the proton, $N_H$ is the column density of the outflow gas, and $c$ is the speed of light in vacuum. The column density of the outflowing gas is typically $N_H = 10^{21}\ cm^{-2}$ for our sample (fig. S7). In this expression, $R_0$, the outflow launching radius, is the only free parameter when we perform the model fitting. For a typical quasar SED, the total luminosity of optical+UV+EUV (extreme ultraviolet) bands is about two-thirds of the bolometric luminosity, i.e., $L_{uv} \simeq (2/3) L_{bol}$, while $L_{bol} = 10^{46.5}$ erg s$^{-1}$ in our sample, typically.

As shown in Fig. 3A, model fitting for our sample results in a best-fit launching radius of $R_0 = 7.8 \pm 0.3$ pc. Although saturation effects are reduced by selecting variable C IV troughs in our sample, the column density may still be underestimated. As a result, we discuss the effects of increasing $N_H$ by a factor of 10. As shown in fig. S8, if we scale up $N_H$ to $10^{22}\ cm^{-2}$, then the observed data cannot be reasonably fitted and the launching radius $R_0$ should be of the order of 1 pc. This result indicates that the outflowing gas with high column density should locate at a small radius. If we keep $v_{max}$ and $\tau_{uv}$ constant, $R_0$ will change from 7.8 to 0.78 pc when $N_H$ increases from $10^{21}$ to $10^{22}\ cm^{-2}$. Considering the fact that the C IV BAL trough is deepest at about 5000 km s$^{-1}$, we adopt an outflow velocity of 5000 km s$^{-1}$ to calculate the mass outflow rate $\dot{M}_{out}$ and the kinetic luminosity $\dot{E}_k$. The locations of an outflow with 5000 km s$^{-1}$ are 8.6 and 0.86 pc for the cases of $N_H = 10^{21}$ and $10^{22}\ cm^{-2}$, respectively, and the time intervals $\Delta T$ that the outflow moves from the launching radius $R_0$ to the locations of outflow with 5000 km s$^{-1}$ are about 350 and 35 years, respectively. As $\dot{M}_{out}$ is proportional to $R^2 N_H/\Delta T$, it is identical in these two cases. Furthermore, because of the same values of $v_{max}$ and $\dot{M}_{out}$, the kinetic luminosity $\dot{E}_k$ also remains the same.

As for J1419, we treat $\tau_{uv}$ as a free parameter and the best-fit launching radius is found to be $R_0 = 10.1 \pm 0.2$ pc (Fig. 3B). The best-fitted $\tau_{UV}$ is $0.043 \pm 0.001$, which is smaller than the mean value $\tau_{UV} = 1$ of the sample. As shown in fig. S5C, the spectrum of J1419 is bluer than the composite spectrum of the sample. The best-fitted extinction for J1419 is $A_V = 0.005 \pm 0.004$ ($\tau_{UV} = 0.05 \pm 0.04$). This value is consistent with the best-fitted value $\tau_{UV} = 0.043 \pm 0.001$ in the dust-driven model. The launching radius is several times of the inner radius of a typical dusty torus, invoking the possibility that BAL outflows likely originate from the torus (Fig. 4). In the dust-driven model mentioned above, the time scale for an outflow to be accelerated from 0 to 10,000 km s$^{-1}$ is of the order of 1000 years (fig. S9), longer than the acceleration time scale of a disk wind by a factor of ~1000. Note that the multi-epoch observations over several years do not yield detectable profile shifts (*50*, *51*) in most of our sample BAL troughs. This rarity appears in line with a typical time scale of outflow acceleration significantly longer than several years.

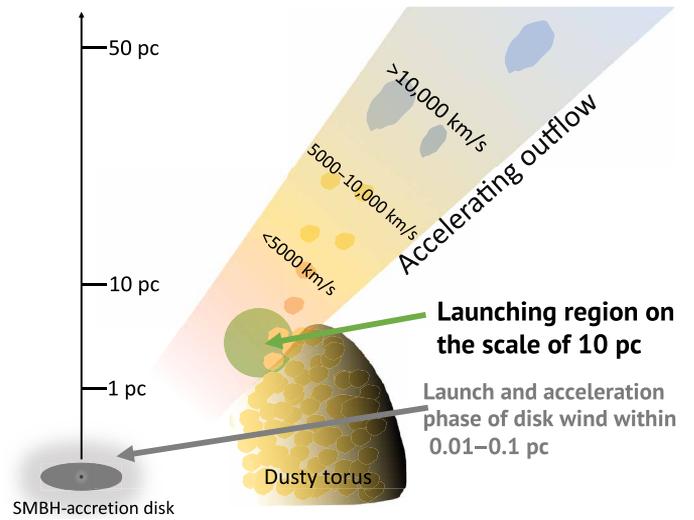

**Fig. 4. The cartoon for the torus-original and dust-driven models for BAL outflow.** The typical launching radius of the torus-original outflow is about 10 pc for the quasar with bolometric luminosity $L_{bol} = 10^{46.5}$ erg s$^{-1}$. The typical launching radius of the disk wind is about 0.01 to 0.1 pc.

The power of the feedback effect is conventionally quantified by the mass outflow rate $\dot{M}_{out}$ and the kinetic luminosity $\dot{E}_k$. The typical overall mass outflow rate for the sample is estimated to be $\dot{M}_{out} = 6.0^{+5.5}_{-0.5}\ M_\odot\ year^{-1}$ (see Methods for details), while the accretion rate is typically $\dot{M}_{acc} = 5.6\ M_\odot\ year^{-1}$ for a quasar with $L_{bol} = 10^{46.5}$ erg s$^{-1}$. Hence, the mass outflow rate is of the same order as the accretion rate. The kinetic-to-bolometric luminosity ratio is therefore $\frac{\dot{E}_k}{L_{bol}} = 1.3^{+1.2}_{-0.1}$%, suggesting that (likely dusty) outflows are sufficiently effective to regulate the growth of the SMBHs and their host galaxies. In conclusion, the presence of dust may lead to efficient coupling between the outflowing gas and the quasar radiation, producing profound effects on the global evolution of quasar-hosting galaxies.

As a rigorous and conservative statement, our sample prefers those unsaturated C IV BAL troughs and the corresponding BAL variations dominated by the variation of the ionizing continuum. Our sample does not represent the diversity of BAL outflow. It is a highly plausible inference that there exists more than one acceleration mechanism for quasar outflows and the acceleration actually occurs on different scales. The C IV BAL may be only diagnostic for a limited range of physical parameters and a limited range of spatial scales, although we think this work will help shed light on the highly perplexing outflow science. Our work confirms that the high-velocity outflow can be accelerated on a scale of tens of pc by the dust-driven model.

## METHODS

### A method to derive the recombination time scale $t_r$

For a typical high-luminosity quasar, we assume that $M_{BH} = 10^9 M_\odot$, $M_i = -25$, the characteristic time scale $\tau = 150$ days, and the value of the structure function $SF(\Delta t) = SF_\infty (1-e^{-|\Delta t|/\tau})^{1/2}$ at infinity $SF_\infty$

= 0.3 mag (*52*, *53*) at 1500 Å. The energy of ionized photons for C III to C IV and C IV to C V is about 47.9 (259 Å) and 64.5 eV (192 Å). Because of the strong absorption in the EUV and soft x-ray, there is not enough observational data to describe its flux variation behavior. According to the empirical formula [equation 5 in (*53*)] from UV/optical data, the relationship between characteristic time scale and wavelength is $\tau \propto \lambda^{2/3}$. Hence, the characteristic time scale at 200 Å deduced from 1500 Å is about 40 days. Assuming that the characteristic time scale is related to the thermal time scale (*52*), the $\tau$ is proportional to wavelength as $\lambda^2$. Then, the $\tau$ of 200 Å is about 2.7 days. To summarize, the $\tau$ of 200 Å may range from a few days to dozens of days. Hence, we adopt 10 days for the simulation and take 3 and 30 days to test the dependence of our method on characteristic time scale $\tau$ in the last subsection. We find that the change of $\tau$ just leads to a tiny shift in the estimation of outflow distance but has no influence on the final conclusion. We generate the light curves from the damped random walk (DRW) model (*54*, *55*) using the Python package astroML (*54*). An example of the light curves of the $\tau$ = 10 days is shown in fig. S1. We pick out three observational epochs from this light curve. The time interval between observational epochs 1 and 2 is $\Delta T_{12}$ = 30 days. The time interval between epochs 1 and 3 is $\Delta T_{13}$ = 175 days. The amplitude of flux variation for a pair of observations is defined as follows: $\Delta L/L = 2(L_2 - L_1)/(L_1 + L_2)$, where $L_1$ and $L_2$ are the fluxes measured at the first and second observations, respectively.

As shown in fig. S1A, the amplitude of flux variation between epochs 1 and 2 is greater than that for epochs 1 and 3, i.e., $(|(\Delta L/L)_{12}| - |(\Delta L/L)_{13}|) \times (\Delta T_{12} - \Delta T_{13}) < 0$. When the recombination time scale $t_r$ is 20 days, the variation of averaged flux over $t_r$ (representing the BAL variation) from epochs 1 to 2 is greater than that from epochs 1 to 3, i.e., $|(\Delta L/L)_{12}|_{mean} > |(\Delta L/L)_{13}|_{mean}$ (the corresponding confidence of BAL variation in observation). When the recombination time scale $t_r$ is 100 days, the variation of average flux from epochs 1 to 2 is smaller than that from epochs 1 to 3, i.e., $|(\Delta L/L)_{12}|_{mean} < |(\Delta L/L)_{13}|_{mean}$. In general, the probability of $|(\Delta L/L)_{12}|_{mean} > |(\Delta L/L)_{13}|_{mean}$ will decrease gradually with $t_r$ (fig. S1B). Therefore, we can measure $t_r$ by the probability of $|(\Delta L/L)_{12}|_{mean} > |(\Delta L/L)_{13}|_{mean}$. For two pairs of observations (such as epochs 1 to 2 and epochs 1 to 3), if the product of the difference of flux variation and the difference of average flux variation $(|(\Delta L/L)_{12}| - |(\Delta L/L)_{13}|) \times (|(\Delta L/L)_{12}|_{mean} - |(\Delta L/L)_{13}|_{mean}) > 0$, then we mark this case as G1 and mark the case of $(|(\Delta L/L)_{12}| - |(\Delta L/L)_{13}|) \times (|(\Delta L/L)_{12}|_{mean} - |(\Delta L/L)_{13}|_{mean}) < 0$ as G2. We define the fraction of G1 as follows: $F_{G1} = N_{G1}/(N_{G1} + N_{G2})$, where $N_{G1}$ and $N_{G2}$ are the numbers of G1 and G2 cases certificated from the observational data, respectively. The G1 fraction $F_{G1}$ decreases when $t_r$ increases. As a result, we can use $F_{G1}$ to constrain $t_r$.

### The sample selection and G1 and G2 certification process

We merge the BAL quasar catalog of SDSS DR7 (*56*) and that of DR12 (*57*). Then, we compare this catalog with SDSS DR16 and select quasars with at least three spectroscopic observations. We adopt a redshift cut (1.9 < $z$ < 4.7) to cover the wavelength range of the C IV λ1549 BAL trough. After the above criteria, we obtain a 14,760 C IV BAL sample and a subsample containing 1559 BAL quasars with at least three spectroscopic observations for each quasar. Note that the few-epoch spectroscopy (FES) program of SDSS chooses BAL quasars [mainly from the sample of (*1*)] with apparent magnitude at *i* band $m_i$ < 19.28 and then obtains at least three observations to study the BAL acceleration or reemergence/disappearance phenomenon [see details in section 3.5 of (*58*)]. The above subsample with at least three observations is brighter than the whole BAL sample for about 0.7 mag.

To ensure detection of BAL variations, we keep only quasars with at least one spectrum with S/N at the SDSS *g* band greater than 10. There are 3418 spectra of 780 quasars matched above criteria. To study different components in a wide velocity range, we select three velocity regions: low velocity (0 to 5000 km s$^{-1}$), medium velocity (5000 to 10,000 km s$^{-1}$), and high velocity (>10,000 km s$^{-1}$). We require the three regions to be detected with BAL variations at the same time.

The demonstration of G1 and G2 certification process is shown in fig. S2. In *n* observations of an object, a, b, and c, which generate three two-pair observations: (ab, ac), (ab, bc), and (ac, bc). Taking the two-pair (ab, bc) as an example, if $\Delta T_{ab} > \Delta T_{bc}$, then we define $\Delta T_{long} = \Delta T_{ab}$ and $\Delta T_{short} = \Delta T_{bc}$. The same notation adopts for $L$ and EW properties. If $|\Delta L/L|_{short} - |\Delta L/L|_{long} > 20\%$ and $|\Delta EW/\sigma_{EW}|_{short} > |\Delta EW/\sigma_{EW}|_{long}$, we classify this case as "G1"; if $|\Delta L/L|_{short} - |\Delta L/L|_{long} > 20\%$ and $|\Delta EW/\sigma_{EW}|_{short} < |\Delta EW/\sigma_{EW}|_{long}$, then mark this case as "G2." An example of the G1 and G2 certification is shown in fig. S3. After the above certification process, we obtain a sample of 137 two-pair spectra from 46 BAL quasars (table S1).

Note that the quasar J1419 (hereafter J1419) at $z$ = 2.145 is excluded from the sample and analyzed separately. J1419 has 72 observations with the S/N level at *g* band greater than 5 in all the epochs. The variation of BAL of J1419 has been found to be driven by the photoionization (*59*–*61*). As shown in fig. S5, there are five different velocity components (from A to E, velocity increasing) according to the visual inspection. After the certification process, there are 3980 two-pair observations left (2559 G1 and 1421 G2). The G1 fraction $F_{G1}$ is 64.3 % ± 0.8% (table S2). The information of other troughs is also shown in table S2.

### Spectral fitting and identification of the C IV BAL variation

To obtain the continuum flux at 1500 Å, we fit the spectra by scaling the unabsorbed template quasar spectra with a double power-law function [equation 1 in (*39*)]. The unabsorbed templates are derived from 38,377 non-BAL quasars from SDSS DR7. The fitting process is the same as in (*39*, *40*, *43*). Please see (*39*, *40*, *43*) for details.

To identify the C IV BAL variation between a pair of spectra, we select the higher S/N spectrum of the pair of spectra as a template to match the other spectrum by rescaling it using the double power-law function. To account for variations of the emission line, we add/subtract a Gaussian to/from the rescaled spectrum. As shown in fig. S3, this rescaled reference matching usually produces a good fit over the continuum and emission-line regions. We then measure C IV BAL variation from the difference between the rescaled higher S/N spectrum and the other spectrum of a pair of spectra. The confidence of the BAL variation is defined as follows: $N_\sigma = \sum |\Delta flux| / \sqrt{\sum \sigma^2}$, where the flux uncertainties ($\sigma = \sqrt{\sigma_{flux_1}^2 + \sigma_{flux_2}^2}$) of the two spectra include the possible systematic uncertainties due to rescaling. After the identification of the C IV BAL variation, we calculate the confidence of BAL variation in the low-velocity, medium-velocity, and high-velocity regions, respectively (fig. S2). The process of identifying C IV BAL variation is the same as in (*39*, *40*, *43*). Please see (*39*, *40*, *43*) for details. The traditional approach of detecting BAL variation is comparing the BAL EWs of different spectra. As a result, the uncertainty of continuum will affect the detection results, especially the shallower absorption at higher velocities. Compared with the traditional approach, our approach of detecting BAL variation avoids the fitting of continuum.

Dividing the spectra by the continuum, we obtain normalized spectrum. As described in (*43*), for a partially obscured absorber, the normalized residual flux in the trough is

$$I(v) = 1 - C(v) + C(v) e^{-\tau(v)} \quad (2)$$

where $C(v)$ is the covering factor and $\tau(v)$ is the optical depth of the ion at velocity $v$ (*62*, *63*). For the resonance doublet C IV 1548.2, 1550.8 Å, their oscillator strengths is $f_{blue} = 0.19$ and $f_{red} = 0.095$, respectively. For the resonance doublet Si IV 1393.8, 1402.8 Å, their oscillator strengths is $f_{blue} = 0.513$ and $f_{red} = 0.255$, respectively. The optical depth ratio of the doublet gives an optical depth ratio of $f_{blue}/f_{red}$ close to 2.

Thus, we will use the doublet components to fit the BAL troughs. As described in (*43*), we simply consider a constant covering factor for the whole BAL trough and allow the optical depth $\tau$ to vary with velocity. According to the partial covering model, we obtain a set of equations of $\tau(v_i)$ as follows

$$\begin{cases} I_{v_1} &= 1 - C + C e^{-\tau(v_1)} \\ &\vdots \\ I_{v_k} &= 1 - C + C e^{-\tau(v_k)} \\ I_{v_{k+1}} &= \left[1 - C + C e^{-\tau(v_{k+1})}\right]\left[1 - C + C e^{-2\tau(v_1)}\right] \\ &\vdots \\ I_{v_{n-k}} &= \left[1 - C + C e^{-\tau(v_{n-k})}\right]\left[1 - C + C e^{-2\tau(v_{n-2k})}\right] \\ I_{v_{n-k+1}} &= 1 - C + C e^{-2\tau(v_{n-2k+1})} \\ &\vdots \\ I_{v_n} &= 1 - C + C e^{-2\tau(v_{n-k})} \end{cases} \quad (3)$$

where $\tau$ is the optical depth of red component. There are $n$ equations in total with $n - k + 1$ unknown variables, where $k = 5$ for a bin of 0.5 Å in wavelength. Since there are more constraints than unknown variables, the equation set has no exact solution. We therefore use the least-squares method to find a set of {$\tau(v_i)$} that best fits these equations. To account for the noise in the flux and the uncertainty of continuum, the low limit of the optical depth $\tau$ is set to −0.1 (corresponding to a normalized flux $I \simeq 1.1$). For most of the troughs, the best-fit results give a reduced $\chi^2$ around 1. An example fit for SDSS J1419 is shown in fig. S5.

We should note that C IV BALs often exhibit nonblack saturation from partial covering and the covering fractions are velocity dependent (*12*, *64*). The covering fraction may be lower at high velocity (*12*), which can lead to BAL at high velocity being often shallower. In this case, the column density at high velocity would be underestimated in our analysis, more so than at low velocity. Furthermore, detection of variability in absorption is much harder when the BAL is shallower and the covering fraction is lower. This may result in an effect similar to that of a low S/N at high velocity. A low detection ability of BAL variation will bring the G1 fraction closer to 50%. Since we observe that the G1 fractions are 67.9, 57.7, and 40.9% for low-, medium-, and high-velocity bins, respectively, meaning that the intrinsic G1 fractions should be >67.9, >57.7, and <40.9%, respectively.

Hence, the difference of intrinsic G1 fractions between the high-velocity bin and the other bins should be more significant. However, we may not fully rule out that the low covering factor at high velocity may still lead to some uncertainties through a path that we do not understand at present. The uncertainty brought by the covering factor is an issue worthy of in-depth consideration in future work given its complexity.

We require the three regions—low-velocity (0 to 5000 km s$^{-1}$), medium-velocity (5000 to 10,000 km s$^{-1}$), and high-velocity regions (>10,000 km s$^{-1}$)—to be detected with BAL variations at the same time. It is hard to imagine outflow components with large velocity differences moving in and out of sight at the same time. Therefore, the BAL variations in our sample are dominated by the variation of the ionizing continuum, and the variation of covering fraction is minimized. As a result, in our sample, the C IV saturation effects are reduced by selecting C IV troughs with simultaneous variation in a wide velocity range. The variation of covering fraction is minimized in our sample.

After the troughs are fitted, the C IV column densities are obtained by integrating the optical depth over the troughs (*65*)

$$N_{ion} = \frac{3.7679 \times 10^{14} \, cm^{-2}}{\lambda f} \int \tau(v) \, dv \quad (4)$$

where $\lambda$ and $f$ are the transition's wavelength and oscillator strength, respectively, and the velocity $v$ is measured in km s$^{-1}$. The distribution of C IV column densities is shown in fig. S7.

### Simulate the G1 curve

We use the time interval and the amplitude of flux variations at 1500 Å of two-pair observations in our sample (table S1) to determine the G1 fraction $F_{G1}$ curve at different recombination time scales $t_r$. Then, the G1 fraction $F_{G1}$ curve and the observed $F_{G1}$ constrain the $t_r$.

We compute the G1 fraction $F_{G1}$ as a function of $t_r$ at values of $\log_{10} t_r$ that are spaced evenly between −0.5 and 3.0 with a step size of 0.1. The detailed process is as follows:

1) Pick up the time interval ($\Delta T_{short}$, $\Delta T_{long}$) and the amplitude of flux variations [($\Delta L/L$)$_{short}$, ($\Delta L/L$)$_{long}$] of a two-pair observation from the sample and then select the data points meeting the requirement of the above four parameters from a random DRW light curve at 1500 Å. We add a Gaussian random error with σ = 6% to the light curve to account for the SDSS spectrophotometric uncertainties (*66*). Calculate the variation of mean flux ($\Delta L/L$)$_{mean, short}$, ($\Delta L/L$)$_{mean, long}$ for a given recombination time scale $t_r$. If the |($\Delta L/L$)|$_{mean, short}$ is greater than |($\Delta L/L$)|$_{mean, long}$, then we will mark it G1, otherwise, G2. Repeat the above process for all the 137 two-pair observations in our sample. Then, a value of the G1 fraction $F_{G1}$ at the given $t_r$ is generated. With more than 1000 repetitions, the mean value of $F_{G1}$ tends to be stable. Adopt the stable value as the final value of $F_{G1}$ at the given $t_r$.

2) Repeat step 1 for the recombination time scale $\log_{10} t_r$ from −0.5 and 3.0 with a step size of 0.1. Then, the G1 fraction $F_{G1}$ ($t_r$) is generated.

As shown in Fig. 1, the $F_{G1}$ ($t_r$) curve decreases when the recombination time scale $t_r$ is increased. The $F_{G1}$ ($t_r$) curve generation process for the J1419 (the black curve in Fig. 2A) is similar to that of the sample.

**The recombination time scale $t_r$ of the C IV line**

The recombination time scale $t_r$ (67) of the C IV line is related to the electron density $n_e$ and the recombination rate α

$$t_r = \left[-f\alpha_{CIV}n_e\left(\frac{n_{CV}}{n_{CIV}} - \frac{\alpha_{CIII}}{\alpha_{CIV}}\right)\right]^{-1} \quad (5)$$

where $f$ is the amplitude of the change in the mean incident ionizing flux between two time periods: One is a period of time $t_r$ ending at the observational point, and another one is a period of time $t_r$ beginning from this observational point (fig. S10). We match the 137 two-pair observations from our sample to the light curve generated by the DRW model to yield the $f$ factor value at different $t_r$. As shown in fig. S10, the typical value of $f$ is about 0.1. Using the Chianti atomic database version 8.0 (68) at a nominal temperature of $2 \times 10^4$ K, we take the recombination rates $\alpha_{CIV} = 5.3 \times 10^{-12}$ cm$^3$ s$^{-1}$ (from C V to C IV) and $\alpha_{CIII} = 2.1 \times 10^{-11}$ cm$^3$ s$^{-1}$ (from C IV to C III). At ionization parameters $\log_{10} U = -0.5, 0$, and 1, the ratio of the number densities of C V to C IV are $\frac{n_{CV}}{n_{CIV}} \approx 30, 100$, and 1000, showing that C IV responds negatively to an increasing ionization parameter, i.e., producing the overionized state for C IV.

**Radiation driving on a dusty cloud**

The equation of motion for radiation driving a dusty cloud can be written as follows

$$\frac{vdv}{dR} = \frac{L_{uv}(1-e^{-\tau_{uv}})}{4\pi R^2 m_p N_H c} - \frac{GM_{BH}}{R^2} - \frac{C_d n_{ISM}}{N_H}v^2 \quad (6)$$

where, $L_{uv} = \lambda L_{Edd}$ is the optical+UV+EUV band luminosity. According to the definition of Eddington luminosity, $L_{Edd} = 4\pi GM_{BH}m_pc/\sigma_T$, so Eq. 6 can be written as follows

$$\frac{vdv}{dR} = \left[\frac{A_d\lambda(1-e^{-\tau_{uv}})}{\tau_{uv}} - 1\right]\frac{GM_{BH}}{R^2} - \frac{C_d n_{ISM}}{N_H}v^2 \quad (7)$$

in which $\tau_{uv} = A_d N_H \sigma_T$ is the optical depth of dusty gas ($\sigma_T = 6.65 \times 10^{-25}$ cm$^{-2}$ is the Thomson cross section). $A_d = \kappa_{uv}/\kappa_T = \tau_{uv}/(N_H\sigma_T)$ is the amplification factor. For our sample quasars, the $A_d = 1500$, which is lower than the typical value $A_d = 2500$ (69) for the average dust-to-gas ratio in the Milky Way. The first term on the right side of the equation is the radiation driving; the second is the gravity of the SMBH, and the last one is the resistance of the interstellar medium. The drag coefficient $C_d$ is close to 1 (70). In our case, $\lambda A_d \gtrsim 100$ and $\tau_{uv} \simeq 1$; therefore, the gravity term can be neglected. When the drag pressure is significantly smaller than the radiation pressure term, the above equation can be simplified as

$$\frac{vdv}{dR} = \frac{L_{uv}(1-e^{-\tau_{uv}})}{4\pi R^2 m_p N_H c} \text{ or } \frac{vdv}{dR} = \frac{A_d\lambda(1-e^{-\tau_{uv}})}{\tau_{uv}}\frac{GM_{BH}}{R^2} \quad (8)$$

**Calculation under three AGN SED shapes**

To determine the hydrogen column density $N_H$, we run a series of photoionization simulations using Cloudy (version c17) (71). Since the gas ionization is insensitive to the electron density at a given ionization parameter, we take a typical electron density $n_e = 10^5$ cm$^{-3}$. The photoionization of an outflow depends on the incident SED. Similar to (43), we compare the photoionization solutions obtained using three different AGN SEDs: UV-soft, MF87, and HE 0238 (fig. S7). The UV-soft SED is used for high-luminosity radio-quiet quasars (72). The MF87 SED (73) is usually used to describe radio-loud quasars, whose most obvious feature is the so-called big blue bump. The HE 0238 SED is used to describe HE 0238-1904 ($z = 0.6309$), which is a radio-quiet quasar (44). In this work, the averaged value of the photoionization solutions for the three SEDs is adopted as our final result.

As mentioned in (44), AGN outflow gases have supersolar metallicities [e.g., Mrk 279, $Z \simeq 2Z_\odot$ (74); SDSS J1512+1119, $1Z_\odot \leq Z \leq 4Z_\odot$ (75); and SDSS J1106+1939, $Z = 4Z_\odot$ (76)]. We therefore adopt a moderate metallicity $Z = 2Z_\odot$ in our calculations. Using the photoionization simulations, we obtain the relations between the hydrogen column density $N_H$ and the C IV column density $N_{CIV}$ at $\log_{10} U = 0$ (see fig. S7). Then, the hydrogen column density $N_H$ of all the outflows can be estimated from these relations. The bolometric luminosities for the three quasar SED types are as follows: $L_{bol} = 4.2\, \lambda_{1500}L_{1500}$ (UV-soft), $L_{bol} = 6.6\, \lambda_{1500}L_{1500}$ (MF87), and $L_{bol} = 4.1\, \lambda_{1500}L_{1500}$ (HE 0238). The bolometric luminosities in the work are based on the averaged result of the three SEDs.

The outflow distance $R$ can be determined as long as the ionization state and density are known. The ionization parameter is defined as follows

$$U = \frac{Q_H}{4\pi R^2 n_H c} \quad (9)$$

where $Q_H = \int_{\nu_0}^{+\infty} \frac{L_\nu}{h\nu} d\nu$ is the source emission rate of hydrogen-ionizing photons and $h\nu_0 = 13.6$ eV is the ionization potential of H$^0$ for ionization out of the 1-s ground state. $c$ is the speed of light in vacuum, and $n_H \approx 0.83 n_e$ is the hydrogen number density. Combining Eqs. 5 and 9, the outflow distance $R$ can be calculated. Considering that part of the incident ionizing photons may be absorbed by the low-velocity component of outflow, we use Cloudy to calculate the transmitted SED. As shown in fig. S7A, the ratio of transmitted to incident ionizing photons $Q_{Hout}/Q_{Hin} \sim 1$, at $\log_{10} U = 0$ and $N_H = 10^{21}$ cm$^{-2}$. This shows that there is very little absorption of ionizing photons by outflowing gas. As a result, we do not need to use the absorbed SED to calculate the distance of the high-velocity component.

Assuming that the outflow is in the form of a thin partial shell ($\Delta R/R \ll 1$), the mass of the outflow $M_{out}$, mass flow rate $\dot{M}_{out}$, and kinetic luminosity $\dot{E}_k$ can be given by (77)

$$M_{out} = 4\pi\Omega R^2 \mu m_p N_H \quad (10)$$

$$\dot{M}_{out} = M_{out}/\Delta T \quad (11)$$

$$\dot{E}_k = \frac{1}{2}\dot{M}_{out}v^2 \quad (12)$$

where μ = 1.4 is the mean atomic mass per proton, $m_p$ is the mass of the proton, and $v$ is the outflow velocity. $\Omega \simeq 0.2$ is the global covering factor that can be deduced by the usual statistical approach for C IV BALs. $\Delta T$ is the time that the outflow moves from the launching radius $R_0$ to current radius $R$. As shown in fig. S4B, the bottom of the C IV BAL trough is at about 5000 km s$^{-1}$, and the left and right boundaries of the half-depth level are 10,000 and 2500 km s$^{-1}$, respectively, according to the composite spectrum. As shown in fig. S9, according to the best-fit dust-driven model, the distances at 2500, 5000, and 10,000 km s$^{-1}$ are 7.8, 8.6, and 14.2 pc, respectively. The times are 150.8, 348.4, and 1036.4 years, respectively. We take the result calculated for velocity 5000 km s$^{-1}$ as the mean value of $\dot{M}_{out}$, and the results calculated for velocities 2500 and 10,000 km s$^{-1}$ as the estimation for the error of $\dot{M}_{out}$. Here, $\dot{M}_{out}$ is the average value for a long time, not the instantaneous value. Hence, $\dot{M}_{out}$ should be the same at different radii $R$. The constant mass outflow rate is $\dot{M}_{out} = 6.0^{+5.5}_{-0.5} M_{\odot}$ year$^{-1}$. The kinetic luminosity $\dot{E}_k$ will increase with radius $R$. The final kinetic-to-bolometric luminosity ratio at infinity is about $\dot{E}_k/L_{bol}(\infty) = 1.3^{+1.2}_{-0.1}$%.

### The ionization parameter $U$ of J1419

To determine the ionization parameter U of the outflow gas in J1419, we run a series of photoionization simulations. As shown in fig. S6A, $N_H - U_H$ plane plot shows the photoionization solution for the five outflow components. The thin and thick curves represent Si IV and C IV, respectively. The solutions of ($\log_{10} N_H(\text{c m}^{-2})$, $\log_{10} U_H$) are (19.4, −2.0), (18.3, −2.1), (18.0, −2.0), (18.4, −1.9), and (18.2, −1.9), respectively. The ionization parameters of all the five outflow components are near $\log_{10} U = -2$. The C IV BAL EW and the monochromatic luminosity at 1500 Å are shown in fig. S6B. The correlation coefficients and $P$ values are marked in each panel. The strong negative correlations between the BAL troughs and the continuum reveal that the variations of BAL troughs are driven by the variation of ionizing continuum. We must point out that the above is just a rough estimate of ionization parameters. In particular, the estimation of ionization parameters of trough A may be terrible because of the possible saturation. However, even if taking out trough A, the recombination time scale $t_r$ is still increasing with the velocity for the other four troughs. Furthermore, because of $R \propto U^{-1/2}$ at a given electron density, the uncertainty of distance is smaller than that of ionization parameter.

### The dependence of our method on characteristic time scale τ

Here, we adopt two other characteristic time scales τ, 3 and 30 days, for the DRW model to test the dependence of our method on characteristic time scale τ. As shown in fig. S11, the G1 fraction indeed increases with the characteristic time scale τ, especially with a short recombination time scale $t_r$. The shift in the long recombination time scale part is small. Take the observed G1 fraction 67.9% of the low velocity (0 to 5000 km s$^{-1}$). The $t_r$ are $10^{0.09}$ and $10^{0.95}$ days at the characteristic time scales τ, 3 and 30 days, respectively. The $t_r$ increases by 0.86 dex when the characteristic time scale τ increases by 1 dex. Because the outflow distance $R \propto t_r^{1/2}$, the derived $R$ will shift by only 0.4 dex (a factor of 2.5) even if the characteristic time scale τ shifts by 1 dex (a factor of 10). As a result, the shift in distance estimation caused by the uncertainty of characteristic time scales has no influence on the final conclusion.

### SUPPLEMENTARY MATERIALS

Supplementary material for this article is available at https://science.org/doi/10.1126/sciadv.abk3291

### REFERENCES AND NOTES

1. R. R. Gibson, L. Jiang, W. N. Brandt, P. B. Hall, Y. Shen, J. Wu, S. F. Anderson, D. P. Schneider, D. V. Berk, S. Gallagher, X. Fan, D. G. York, A catalog of broad absorption line quasars in Sloan Digital Sky Survey data release 5. *Astrophys. J.* **692**, 758–777 (2009).
2. J. T. Allen, P. C. Hewett, N. Maddox, G. T. Richards, V. Belokurov, A strong redshift dependence of the broad absorption line quasar fraction. *Mon. Not. R. Astron. Soc.* **410**, 860–884 (2011).
3. I. G. McCarthy, J. Schaye, T. J. Ponman, R. G. Bower, C. M. Booth, C. D. Vecchia, R. A. Crain, V. Springel, T. Theuns, R. P. C. Wiersma, The case for agn feedback in galaxy groups. *Mon. Not. R. Astron. Soc.* **406**, 822–839 (2010).
4. N. Soker, A moderate cooling flow phase at galaxy formation. *Mon. Not. R. Astron. Soc.* **407**, 2355–2361 (2010).
5. A. C. Fabian, Observational evidence of active galactic nuclei feedback. *Annu. Rev. Astron. Astrophys.* **50**, 455–489 (2012).
6. C.-A. Faucher-Giguère, E. Quataert, N. Murray, A physical model of felobals: Implications for quasar feedback. *Mon. Not. R. Astron. Soc.* **420**, 1347–1354 (2012).
7. E. Choi, T. Naab, J. P. Ostriker, P. H. Johansson, B. P. Moster, Consequences of mechanical and radiative feedback from black holes in disc galaxy mergers. *Mon. Not. R. Astron. Soc.* **442**, 440–453 (2014).
8. F. W. Hamann, T. A. Barlow, F. C. Chaffee, C. B. Foltz, R. J. Weymann, High-resolution keck spectra of the associated absorption lines in 3c 191. *Astrophys. J.* **550**, 142–152 (2001).
9. D. M. Capellupo, F. Hamann, T. A. Barlow, A variable PV broad absorption line and quasar outflow energetics. *Mon. Not. R. Astron. Soc.* **444**, 1893–1900 (2014).
10. A. B. Lucy, K. M. Leighly, D. M. Terndrup, M. Dietrich, S. C. Gallagher, Tracing the outflow of a z = 0.334 FeLoBAL: New constraints from low-ionization absorbers in FBQS J1151+3822. *Astrophys. J.* **783**, 58–78 (2014).
11. N. Arav, G. Liu, X. Xu, J. Stidham, C. Benn, C. Chamberlain, Evidence that 50% of balqso outflows are situated at least 100 pc from the central source. *Astrophys. J.* **857**, 60–78 (2018).
12. K. M. Leighly, D. M. Terndrup, S. C. Gallagher, G. T. Richards, M. Dietrich, The z = 0.54 loBAL quasar SDSS j085053. 12+ 445122.5. I. Spectral synthesis analysis reveals a massive outflow. *Astrophys. J.* **866**, 7–27 (2018).
13. F. Hamann, H. Herbst, I. Paris, D. Capellupo, On the structure and energetics of quasar broad absorption-line outflows. *Mon. Not. R. Astron. Soc.* **483**, 1808–1828 (2019).
14. S. Veilleux, G. Cecil, J. Bland-Hawthorn, Galactic winds. *Annu. Rev. Astron. Astrophys.* **43**, 769–826 (2005).
15. T. C. Fischer, D. M. Crenshaw, S. B. Kraemer, H. R. Schmitt, Determining inclinations of active galactic nuclei via their narrow-line region kinematics. I. observational results. *Astrophys. J. Suppl. Ser.* **209**, 1–32 (2013).
16. C. Cicone, M. Brusa, C. R. Almeida, G. Cresci, B. Husemann, V. Mainieri, The largely unconstrained multiphase nature of outflows in AGN host galaxies. *Nat. Astron.* **2**, 176–178 (2018).
17. C. M. Harrison, T. Costa, C. N. Tadhunter, A. Flütsch, D. Kakkad, M. Perna, G. Vietri, AGN outflows and feedback twenty years on. *Nat. Astron.* **2**, 198–205 (2018).
18. D. S. N. Rupke, A review of recent observations of galactic winds driven by star formation. *Galaxies* **6**, 138–161 (2018).
19. S. Veilleux, R. Maiolino, A. D. Bolatto, S. Aalto, Cool outflows in galaxies and their implications. *Astron. Astrophys. Rev.* **28**, 1–173 (2020).
20. N. Murray, J. Chiang, S. A. Grossman, G. M. Voit, Accretion disk winds from active galactic nuclei. *Astrophys. J.* **451**, 498–509 (1995).
21. D. Proga, J. M. Stone, T. R. Kallman, Dynamics of line-driven disk winds in active galactic nuclei. *Astrophys. J.* **543**, 686–696 (2000).
22. A. Konigl, J. F. Kartje, Disk-driven hydromagnetic winds as a key ingredient of active galactic nuclei unification schemes. *Astrophys. J.* **434**, 446–467 (1994).
23. N. Scoville, C. Norman, Stellar contrails in quasi-stellar objects: The origin of broad absorption lines. *Astrophys. J.* **451**, 510–524 (1995).
24. M. Elitzur, I. Shlosman, The agn-obscuring torus: The end of the "doughnut paradigm"? *Astrophys. J. Lett.* **648**, L101–L104 (2006).
25. S. C. Gallagher, J. E. Everett, M. M. Abado, S. K. Keating, Investigating the structure of the windy torus in quasars. *Mon. Not. R. Astron. Soc.* **451**, 2991–3000 (2015).
26. M. Brotherton, H. D. Tran, R. Becker, M. D. Gregg, S. Laurent-Muehleisen, R. White, Composite spectra from the FIRST Bright Quasar Survey. *Astrophys. J.* **546**, 775–781 (2001).
27. T. A. Reichard, G. T. Richards, P. B. Hall, D. P. Schneider, D. E. V. Berk, X. Fan, D. G. York, G. Knapp, J. Brinkmann, Continuum and emission-line properties of broad absorption line quasars. *Astron. J.* **126**, 2594–2607 (2003).
28. A. Baskin, A. Laor, F. Hamann, The average absorption properties of broad absorption line quasars at 800 < $\lambda_{rest}$ < 3000 Å, and the underlying physical parameters. *Mon. Not. R. Astron. Soc.* **432**, 1525–1543 (2013).
29. C. M. Krawczyk, G. T. Richards, S. C. Gallagher, K. M. Leighly, P. C. Hewett, N. P. Ross, P. B. Hall, Mining for dust in type 1 quasars. *Astron. J.* **149**, 203–222 (2015).
30. D. M. Capellupo, F. Hamann, J. C. Shields, P. Rodríguez Hidalgo, T. A. Barlow, Variability in quasar broad absorption line outflows–I. Trends in the short-term versus long-term data. *Mon. Not. R. Astron. Soc.* **413**, 908–920 (2011).
31. N. Filiz Ak, W. Brandt, P. Hall, D. Schneider, S. Anderson, F. Hamann, B. Lundgren, A. D. Myers, I. Pâris, P. Petitjean, N. P. Ross, Y. Shen, D. York, Broad absorption line variability on multi-year timescales in a large quasar sample. *Astrophys. J.* **777**, 168–196 (2013).
32. N. F. Ak, W. Brandt, P. Hall, D. Schneider, J. Trump, S. Anderson, F. Hamann, A. D. Myers, I. Pâris, P. Petitjean, N. Ross, Y. Shen, D. York, The dependence of C IV broad absorption line properties on accompanying Si IV and Al III absorption: Relating quasar-wind ionization levels, kinematics, and column densities. *Astrophys. J.* **791**, 88–109 (2014).

33. S. McGraw, W. Brandt, C. Grier, N. Filiz Ak, P. Hall, D. Schneider, S. Anderson, P. Green, T. Hutchinson, C. Macleod, M. Vivek, Broad absorption line disappearance and emergence using multiple-epoch spectroscopy from the Sloan Digital Sky Survey. *Mon. Not. R. Astron. Soc.* **469**, 3163–3184 (2017).

34. W. Yi, W. N. Brandt, P. B. Hall, M. Vivek, C. J. Grier, N. F. Ak, D. P. Schneider, S. M. McGraw, Variability of low-ionization broad absorption-line quasars based on multi-epoch spectra from the Sloan Digital Sky Survey. *Astrophys. J. Suppl. Ser.* **242**, 28–50 (2019).

35. S. Muzahid, R. Srianand, J. Charlton, M. Eracleous, On the covering fraction variability in an EUV mini-BAL outflow from PG 1206+ 459. *Mon. Not. R. Astron. Soc.* **457**, 2665–2674 (2016).

36. C. Grier, P. Hall, W. Brandt, J. Trump, Y. Shen, M. Vivek, N. F. Ak, Y. Chen, K. Dawson, K. Denney, P. J. Green, L. Jiang, C. S. Kochanek, I. D. McGreer, I. Pâris, B. M. Peterson, D. P. Schneider, C. Tao, W. M. Wood-Vasey, D. Bizyaev, J. Ge, K. Kinemuchi, D. Oravetz, K. Pan, A. Simmons, The Sloan Digital Sky Survey reverberation mapping project: Rapid CIV broad absorption line variability. *Astrophys. J.* **806**, 111–125 (2015).

37. F. Saturni, D. Trevese, F. Vagnetti, M. Perna, M. Dadina, A multi-epoch spectroscopic study of the BAL quasar APM 08279+5255. II. Emission- and absorption-line variability time lags. *Astron. Astrophys.* **587**, A43–A58 (2016).

38. L. Sun, H. Zhou, T. Ji, P. Jiang, B. Liu, W. Liu, X. Pan, X. Shi, J. Wang, T. Wang, C. Yang, S. Zhang, L. P. Miller, Photoionization-driven absorption-line variability in balmer absorption line quasar LBQS 1206+1052. *Astrophys. J.* **838**, 88–109 (2017).

39. T. Wang, C. Yang, H. Wang, G. Ferland, Evidence for photoionization-driven broad absorption line variability. *Astrophys. J.* **814**, 150–165 (2015).

40. Z. He, T. Wang, H. Zhou, W. Bian, G. Liu, C. Yang, L. Dou, L. Sun, W. Bian, Variation of ionizing continuum: The main driver of broad absorption line variability. *Astrophys. J. Suppl. Ser.* **229**, 22–30 (2017).

41. T. A. Barlow, V. T. Junkkarinen, E. M. Burbidge, R. J. Weymann, S. L. Morris, K. T. Korista, Broad absorption-line time variability in the QSO CSO 203. *Astrophys. J.* **397**, 81–87 (1992).

42. J. H. Krolik, G. A. Kriss, Observable properties of x-ray-heated winds in active galactic nuclei: Warm reflectors and warm absorbers. *Astrophys. J.* **447**, 512–525 (1995).

43. Z. He, T. Wang, G. Liu, H. Wang, W. Bian, K. Tchernyshyov, G. Mou, Y. Xu, H. Zhou, R. Green, J. Xu, The properties of broad absorption line outflows based on a large sample of quasars. *Nat. Astron* **3**, 265–271 (2019).

44. N. Arav, B. Borguet, C. Chamberlain, D. Edmonds, C. Danforth, Quasar outflows and AGN feedback in the extreme UV: HST/COS observations of he 0238-1904. *Mon. Not. R. Astron. Soc.* **436**, 3286–3305 (2013).

45. S. Koshida, T. Minezaki, Y. Yoshii, Y. Kobayashi, Y. Sakata, S. Sugawara, K. Enya, M. Suganuma, H. Tomita, T. Aoki, B. A. Peterson, Reverberation measurements of the inner radius of the dust torus in 17 seyfert galaxies. *Astrophys. J.* **788**, 159–179 (2014).

46. S. Zhang, H. Wang, T. Wang, F. Xing, K. Zhang, H. Zhou, P. Jiang, Outflow and hot dust emission in broad absorption line quasars. *Astrophys. J.* **786**, 42–56 (2014).

47. K. D. Gordon, G. C. Clayton, K. Misselt, A. U. Landolt, M. J. Wolff, A quantitative comparison of the small Magellanic cloud, large Magellanic cloud, and milky way ultraviolet to near-infrared extinction curves. *Astrophys. J.* **594**, 279–293 (2003).

48. C. M. Gaskell, J. J. Gill, J. Singh, Attenuation from the optical to the extreme ultraviolet by dust associated with broad absorption line quasars: The driving force for outflows. arXiv:1611.03733 [astro-ph.GA] (2016).

49. J. C. Weingartner, B. T. Draine, Dust grain-size distributions and extinction in the milky way, large Magellanic cloud, and small Magellanic cloud. *Astrophys. J.* **548**, 296–309 (2001).

50. P. B. Hall, S. I. Sadavoy, D. Hutsemekers, J. E. Everett, A. Rafiee, Acceleration and substructure constraints in a quasar outflow. *Astrophys. J.* **665**, 174–186 (2007).

51. C. Grier, W. Brandt, P. Hall, J. Trump, N. F. Ak, S. Anderson, P. J. Green, D. Schneider, M. Sun, M. Vivek, T. G. Beatty, J. R. Brownstein, A. Roman-Lopes, C IV broad absorption line acceleration in Sloan Digital Sky Survey quasars. *Astrophys. J.* **824**, 130–151 (2016).

52. C. L. MacLeod, Ž. Ivezić, C. Kochanek, S. Kozłowski, B. Kelly, E. Bullock, A. Kimball, B. Sesar, D. Westman, K. Brooks, R. Gibson, A. C. Becker, W. H. de Vries, Modeling the time variability of SDSS stripe 82 quasars as a damped random walk. *Astrophys. J.* **721**, 1014–1033 (2010).

53. H. Guo, J. Wang, Z. Cai, M. Sun, How far is quasar UV/optical variability from a damped random walk at low frequency? *Astrophys. J.* **847**, 132–141 (2017).

54. B. C. Kelly, J. Bechtold, A. Siemiginowska, Are the variations in quasar optical flux driven by thermal fluctuations? *Astrophys. J.* **698**, 895–910 (2009).

55. S. Kozłowski, C. S. Kochanek, A. Udalski, I. Soszyński, M. Szymański, M. Kubiak, G. Pietrzyński, O. Szewczyk, K. Ulaczyk, R. Poleski; The OGLE Collaboration, Quantifying quasar variability as part of a general approach to classifying continuously varying sources. *Astrophys. J.* **708**, 927–945 (2009).

56. Y. Shen, G. T. Richards, M. A. Strauss, P. B. Hall, D. P. Schneider, S. Snedden, D. Bizyaev, H. Brewington, V. Malanushenko, E. Malanushenko, M. Malanushenko, D. Oravetz, K. Pan, A. Simmons, A catalog of quasar properties from Sloan Digital Sky Survey data release 7. *Astrophys. J. Suppl. Ser.* **194**, 45–65 (2011).

57. I. Pâris, P. Petitjean, N. P. Ross, A. D. Myers, É. Aubourg, A. Streblyanska, S. Bailey, É. Armengaud, N. Palanque-Delabrouille, C. Yèche, F. Hamann, M. A. Strauss, F. D. Albareti, J. Bovy, D. Bizyaev, W. Niel Brandt, M. Brusa, J. Buchner, J. Comparat, R. A. C. Croft, T. Dwelly, X. Fan, A. Font-Ribera, A. Ge, A. Georgakakis, P. B. Hall, L. Jian, K. Kinemuchi, E. Malanushenko, V. Malanushenko, R. G. McMahon, M. Menzel, A. Merloni, K. Nandra, P. Noterdaeme, D. Oravetz, K. Pan, M. M. Pieri, F. Prada, M. Salvato, D. J. Schlegel, D. P. Schneider, A. Simmons, M. Viel, D. H. Weinberg, L. Zhu, The Sloan Digital Sky Survey quasar catalog: Twelfth data release. *Astron. Astrophys.* **597**, A79–A103 (2017).

58. C. L. MacLeod, P. J. Green, S. F. Anderson, M. Eracleous, J. J. Ruan, J. Runnoe, W. N. Brandt, C. Badenes, J. Greene, E. Morganson, S. J. Schmidt, A. Schwope, Y. Shen, R. Amaro, A. Lebleu, N. Filiz Ak, C. J. Grier, D. Hoover, S. M. McGraw, K. Dawson, P. B. Hall, S. L. Hawley, V. Mariappan, A. D. Myers, I. Paris, D. P. Schneider, K. G. Stassun, M. A. Bershady, M. R. Blanton, J. Seo, J. Tinker, J. G. Fernandez-Trincado, D. J. Chambers, N. Kaiser, K. Kudritzki, E. Magnier, N. Metcalfe, C. Z. Waters, The time-domain spectroscopic survey: Target selection for repeat spectroscopy. *Astron. J.* **155**, 6–22 (2018).

59. Z. Hemler, C. Grier, W. Brandt, P. Hall, K. Horne, Y. Shen, J. Trump, D. Schneider, M. Vivek, D. Bizyaev, A. Oravetz, D. Oravetz, K. Pan, The Sloan Digital Sky Survey reverberation mapping project: Systematic investigations of short-timescale C IV broad absorption line variability. *Astrophys. J.* **872**, 21–42 (2019).

60. W.-J. Lu, Y.-R. Lin, Correlations between the variation of the ionizing continuum and broad absorption lines in individual quasars. *Astrophys. J.* **883**, 30–35 (2019).

61. Q. Zhao, Z. He, G. Liu, T. Wang, H. Guo, L. Shen, G. Mou, A sharp rise in the detection rate of broad absorption line variations in a quasar SDSS J141955.26+522741.1. *Astrophys. J. Lett.* **906**, L8–L13 (2021).

62. N. Arav, K. T. Korista, M. De Kool, V. T. Junkkarinen, M. C. Begelman, Hubble space telescope observations of the broad absorption line quasarpg 0946+301. *Astrophys. J.* **516**, 27–46 (1999).

63. P. B. Hall, S. F. Anderson, M. A. Strauss, D. G. York, G. T. Richards, X. Fan, G. Knapp, D. P. Schneider, D. E. V. Berk, T. Geballe, A. E. Bauer, R. H. Becker, M. Davis, H. Rix, R. C. Nichol, N. A. Bahcall, J. Brinkmann, R. Brunner, A. J. Connolly, I. Csabai, M. Doi, M. Fukugita, J. E. Gunn, Z. Haiman, M. Harvanek, T. M. Heckman, G. S. Hennessy, N. Inada, Z. Ivezic, D. Johnston, S. Kleinman, J. H. Krolik, J. Krzesinski, P. Z. Kunszt, D. Q. Lamb, D. C. Long, R. H. Lupton, G. Miknaitis, J. A. Munn, V. K. Narayanan, E. Neilsen, P. R. Newman, A. Nitta, S. Okamura, L. Pentericci, J. R. Pier, D. J. Schlegel, S. Snedden, A. S. Szalay, A. R. Thakar, Z. Tsvetanov, R. L. White, W. Zheng, Unusual broad absorption line quasars from the Sloan Digital Sky Survey. *Astrophys. J. Suppl. Ser.* **141**, 267–309 (2002).

64. K. M. Leighly, D. M. Terndrup, A. B. Lucy, H. Choi, S. C. Gallagher, G. T. Richards, M. Dietrich, C. Raney, The z = 0.54 LoBAL quasar SDSS J085053.12+445122.5. II. The nature of partial covering in the broad-absorption-line outflow. *Astrophys. J.* **879**, 27–58 (2019).

65. B. D. Savage, K. R. Sembach, The analysis of apparent optical depth profiles for interstellar absorption lines. *Astrophys. J.* **379**, 245–259 (1991).

66. Y. Shen, W. Brandt, K. S. Dawson, P. B. Hall, I. D. McGreer, S. F. Anderson, Y. Chen, K. D. Denney, S. Eftekharzadeh, X. Fan, Y. Gao, P. J. Green, J. E. Greene, L. C. Ho, K. Horne, L. Jiang, B. C. Kelly, K. Kinemuchi, C. S. Kochanek, I. Pâris, C. M. Peters, B. M. Peterson, P. Petitjean, K. Ponder, G. T. Richards, D. P. Schneider, A. Seth, R. N. Smith, M. A. Strauss, C. Tao, J. R. Trump, W. M. Wood-Vasey, Y. Zu, D. J. Eisenstein, K. Pan, D. Bizyaev, V. Malanushenko, E. Malanushenko, D. Oravetz, The Sloan Digital Sky Survey reverberation mapping project: Technical overview. *Astrophys. J. Suppl. Ser.* **216**, 4–28 (2014).

67. N. Arav, D. Edmonds, B. Borguet, G. Kriss, J. Kaastra, E. Behar, S. Bianchi, M. Cappi, E. Costantini, R. Detmers, J. Ebrero, M. Mehdipour, S. Paltani, P. Petrucci, C. Pinto, G. Ponti, K. Steenbrugge, C. de Vries, Multiwavelength campaign on Mrk 509 X. Lower limit on the distance of the absorber from HST COS and STIS spectroscopy. *Astron. Astrophys.* **544**, A33–A45 (2012).

68. G. Del Zanna, K. Dere, P. Young, E. Landi, H. Mason, Chianti–An atomic database for emission lines. Version 8. *Astron. Astrophys.* **582**, A56–A67 (2015).

69. W. Ishibashi, A. C. Fabian, Agn feedback: Galactic-scale outflows driven by radiation pressure on dust. *Mon. Not. R. Astron. Soc.* **451**, 93–102 (2015).

70. L. J. Dursi, C. Pfrommer, Draping of cluster magnetic fields over bullets and bubbles—Morphology and dynamic effects. *Astrophys. J.* **677**, 993–1018 (2008).

71. G. Ferland, M. Chatzikos, F. Guzmán, M. Lykins, P. Van Hoof, R. Williams, N. Abel, F. Badnell, F. Keenan, R. Porter, P. C. Stancil, The 2017 release of cloudy. *Rev. Mex. Astron. Astrofis.* **53**, 385–438 (2017).

72. J. P. Dunn, M. Bautista, N. Arav, M. Moe, K. Korista, E. Costantini, C. Benn, S. Ellison, D. Edmonds, The quasar outflow contribution to AGN feedback: VLT measurements of SDSS J0318-0600. *Astrophys. J.* **709**, 611–631 (2010).

73. W. G. Mathews, G. J. Ferland, What heats the hot phase in active nuclei? *Astrophys. J.* **323**, 456–467 (1987).
74. N. Arav, J. R. Gabel, K. T. Korista, J. S. Kaastra, G. A. Kriss, E. Behar, E. Costantini, C. M. Gaskell, A. Laor, C. N. Kodituwakku, D. Proga, M. Sako, J. E. Scott, K. C. Steenbrugge, Chemical abundances in an AGN environment: X-ray/UV campaign on the markarian 279 outflow. *Astrophys. J.* **658**, 829–839 (2007).
75. B. C. Borguet, D. Edmonds, N. Arav, C. Benn, C. Chamberlain, Bal phosphorus abundance and evidence for immense ionic column densities in quasar outflows: VLT/x-shooter observations of quasar SDSS J1512+1119. *Astrophys. J.* **758**, 69–78 (2012).
76. B. C. Borguet, N. Arav, D. Edmonds, C. Chamberlain, C. Benn, Major contributor to AGN feedback: VLT X-shooter observations of SIV BAL QSO outflows. *Astrophys. J.* **762**, 49–61 (2013).
77. B. C. Borguet, D. Edmonds, N. Arav, J. Dunn, G. A. Kriss, A 10 kpc scale seyfert galaxy outflow: *HST*/cos observations of iras F22456–5125. *Astrophys. J.* **751**, 107–121 (2012).

**Acknowledgments:** Funding for the SSDS-IV has been provided by the Alfred P. Sloan Foundation, the U.S. Department of Energy Office of Science, and the Participating Institutions. SDSS-IV acknowledges support and resources from the Center for High Performance Computing at the University of Utah. The SDSS website is www.sdss.org. SDSS-IV is managed by the Astrophysical Research Consortium for the Participating Institutions of the SDSS Collaboration including the Brazilian Participation Group, the Carnegie Institution for Science, Carnegie Mellon University, Center for Astrophysics—Harvard & Smithsonian, the Chilean Participation Group, the French Participation Group, Instituto de Astrofísica de Canarias, Johns Hopkins University, Kavli Institute for the Physics and Mathematics of the Universe (IPMU)/University of Tokyo, the Korean Participation Group, Lawrence Berkeley National Laboratory, Leibniz Institut für Astrophysik Potsdam (AIP), Max-Planck-Institut für Astronomie (MPIA Heidelberg), Max-Planck-Institut für Astrophysik (MPA Garching), Max-Planck-Institut für Extraterrestrische Physik (MPE), National Astronomical Observatories of China, New Mexico State University, New York University, University of Notre Dame, Observatário Nacional/MCTI, The Ohio State University, Pennsylvania State University, Shanghai Astronomical Observatory, United Kingdom Participation Group, Universidad Nacional Autónoma de México, University of Arizona, University of Colorado Boulder, University of Oxford, University of Portsmouth, University of Utah, University of Virginia, University of Washington, University of Wisconsin, Vanderbilt University, and Yale University. **Funding:** Z.H. is supported by National Natural Science Foundation of China (nos. 11903031, 12192220, and 12192221) and USTC Research Funds of the Double First-Class Initiative YD 3440002001. T.W. is supported by National Natural Science Foundation of China (no. 11833007). G.L. acknowledges the research grants from the China Manned Space Project (nos. CMS-CSST-2021-A06 and CMS-CSST-2021-A07), the National Natural Science Foundation of China (no. 11421303), and the Fundamental Research Funds for the Central Universities. G.M. was supported by the National Natural Science Foundation of China (no. 11703022) and the Fundamental Research Funds for the Central Universities (no. 2042019kf0040). W.B. was supported by the National Natural Science Foundation of China (no. 11973029). L.C.H. was supported by the National Science Foundation of China (nos. 11721303 and 11991052) and the National Key R&D Program of China (no. 2016YFA0400702). H.G. acknowledges the NSF grant AST-1907290. M.S. is supported by NSFC-11973002. **Author contributions:** Z.H. presented the idea, made the calculations, and wrote the manuscript. G.L., T.W., and G.M. discussed the calculations. G.M. gave comments on the dust-driven model. G.L. and R.G. revised the manuscript. All authors discussed and gave comments on the contents of the paper. **Competing interests:** The authors declare that they have no competing interests. **Data and materials availability:** All data needed to evaluate the conclusions in the paper are present in the paper and/or the Supplementary Materials.

Submitted 6 July 2021
Accepted 20 December 2021
Published 11 February 2022
10.1126/sciadv.abk3291